\RequirePackage[2020-02-02]{latexrelease}
\documentclass[twocolumn,english,aps,prl,10pt,superscriptaddress]{revtex4}
\usepackage[T1]{fontenc}
\usepackage{amsmath,graphicx,amssymb,epsfig,dsfont,mathrsfs,color,amsbsy}

\begin{document}

\title{Broadband directional thermal emission with anisothermal microsources}

\author{F. Herz}
\affiliation{Laboratoire Charles Fabry, UMR 8501, Institut d'Optique, CNRS, Universit\'{e} Paris-Saclay, 2 Avenue Augustin Fresnel, 91127 Palaiseau Cedex, France}
\author{R. Messina}
\affiliation{Laboratoire Charles Fabry, UMR 8501, Institut d'Optique, CNRS, Universit\'{e} Paris-Saclay, 2 Avenue Augustin Fresnel, 91127 Palaiseau Cedex, France}
\author{P. Ben-Abdallah}
\email{pba@institutoptique.fr} 
\affiliation{Laboratoire Charles Fabry, UMR 8501, Institut d'Optique, CNRS, Universit\'{e} Paris-Saclay, 2 Avenue Augustin Fresnel, 91127 Palaiseau Cedex, France}

\date{\today}


\begin{abstract}
Thermal emission is naturally spatially incoherent and lacks directionality. Here, we demonstrate that by precisely controlling the spatial temperature distribution within a solid, directional thermal emission can be achieved across a broad spectral range. These anisothermal sources open new avenues for manipulating radiative heat flux at the microscale and hold promise for applications in thermal management and energy conversion, enabling more efficient and targeted thermal control.
\end{abstract}

\maketitle
\subsection*{I. INTRODUCTION}
Thermally generated light is a fundamental phenomenon resulting from the agitation of partial charges in matter due to the current fluctuations induced by the local temperature. Its control is crucial for numerous technologies like energy conversion, imaging or sensing. Traditional thermal emitters are incoherent~\cite{Planck,Kirchoff} and lack directionality due to the random nature of the emission process. Various strategies, such as surface-polariton gratings~\cite{Hesketh86,Hesketh88,Kreiter,Greffet}, metasurfaces~\cite{Wang}, cavities~\cite{Kollyukh,PBA_JOSA,Celanovic}, and photonic crystals~\cite{Lee,Battula} have demonstrated angular control of thermal emission, but this control is limited to narrow bandwidths. Achieving broadband thermal emission confined to specific directions is a much more complex problem since it presents several fundamental challenges. Indeed, thermal emission spans a wide range of wavelengths and controlling the directionality across this broad spectrum is inherently difficult since matter structuring is generally well adapted to the control of specific wavelengths. However, recently several mechanisms have been proposed to achieve a broadband directional thermal emission. Among these is the perfect absorption at the Brewster angle in hyperbolic materials of type II over a large spectral domain~\cite{Barbillon}. However, these sources emit only p-polarized light, which intrinsically limits their flux. A second mechanism is the opening of angular band gaps~\cite{Hamam,Shen} with stacks of anisotropic photonic crystals. In such systems light can only propagate in specific directions of space, all other directions being forbidden due to destructive interferences. Finally, gradient epsilon-near-zero materials have been proposed~\cite{Xu} to achieve broadband directional control. These media support leaky p-polarized Berreman modes able to couple to free-space modes at specific angles across a wide frequency range. In practice, the application of these directive sources is impaired by challenges such as fabrication complexity and material losses. Moreover, they do not allow an active change of the direction of emission once the source is designed. In the present Letter, we introduce the concept of anisothermal sources, which are sources with spatially variable temperatures profiles. By analyzing their emission, we demonstrate that directivity of radiative heat flux can be achieved through a precise control of their temperatures, showing in this way the tunability potential of these thermal antennas. 

\subsection*{II. THEORY}
\subsubsection*{A. Heat flux radiated by an anisothermal source}
To start, we consider a finite-size object of volume $\mathcal{V}$ and of frequency-dependent permittivity $\epsilon(\omega)$ immersed in a thermal bath at temperature $T_\mathrm{b}$. We also assume that the temperature distribution $T(\mathbf{r})$ within this object is non-uniform. The thermal light radiated by this object toward its surrounding environment can be calculated using the fluctuational-electrodynamics theoretical framework~\cite{Rytov}. 

This calculation can be done easily by using the discrete-dipole approximation method~\cite{pba2011, Edalatpour, Cuevas} which consists in discretizing the solid into a finite number $N$ of small polarizable fluctuating dipoles arranged on a grid. Each dipole represents a small volume element of the material and responds to the local electromagnetic field. Those dipoles interact with each other and with the external bath. The local electric and magnetic fields $ \mathbf{E}$ and $\mathbf{H}$ in the surrounding environment are linearly related to fluctuating dipolar moments $\mathbf{p}_{j}^\mathrm{(fl)}$ and to the fields $\mathbf{E}_\mathrm{b}$ and $\mathbf{H}_\mathrm{b}$ radiated by the bath as follows~\cite{pba2011,messina2013,biehs1}
\begin{equation}\begin{split}
 \mathbf{E}(\mathbf{r},\omega)&=\mathds{B}^{\mathrm{(E)}}(\mathbf{r},\omega)\mathbf{E}_\mathrm{b}(\mathbf{r},\omega),\\
 &\,+\mu_{0}\omega^{2}\sum_{j=1}^N\mathds{G}^{\mathrm{(EE)}}(\mathbf{r},\mathbf{r}_j,\omega)\mathbf{p}_{j}^\mathrm{(fl)}(\omega),
 \label{Eq:field_fluc}
\end{split}\end{equation}
and
\begin{equation}\begin{split}
 \mathbf{H}(\mathbf{r},\omega)&=\mathds{B}^{\mathrm{(H)}}(\mathbf{r},\omega)\mathbf{H}_\mathrm{b}(\mathbf{r},\omega)\\
 &\,+\mu_0\omega^2\sum_{j=1}^N\mathds{G}^{\mathrm{(HE)}}(\mathbf{r},\mathbf{r}_j,\omega)\mathbf{p}_{j}^\mathrm{(fl)}(\omega).
 \label{Eq:mag_fluc}
\end{split}\end{equation}
Here, $\mathds{G}^{\mathrm{(EE)}}$ and $\mathds{G}^{\mathrm{(HE)}}$, representing the full electric and magnetic dyadic Green tensors at frequency $\omega$, as well as the matrices $\mathds{B}^{\mathrm{(E)}}$ and $\mathds{B}^{\mathrm{(H)}}$ take into account all many-body interactions. These matrices can be deduced by solving a self-consistent linear system as detailed in Ref.~\onlinecite{messina2013}. 

\begin{figure}
	\centering
	\includegraphics[width=0.45\textwidth]{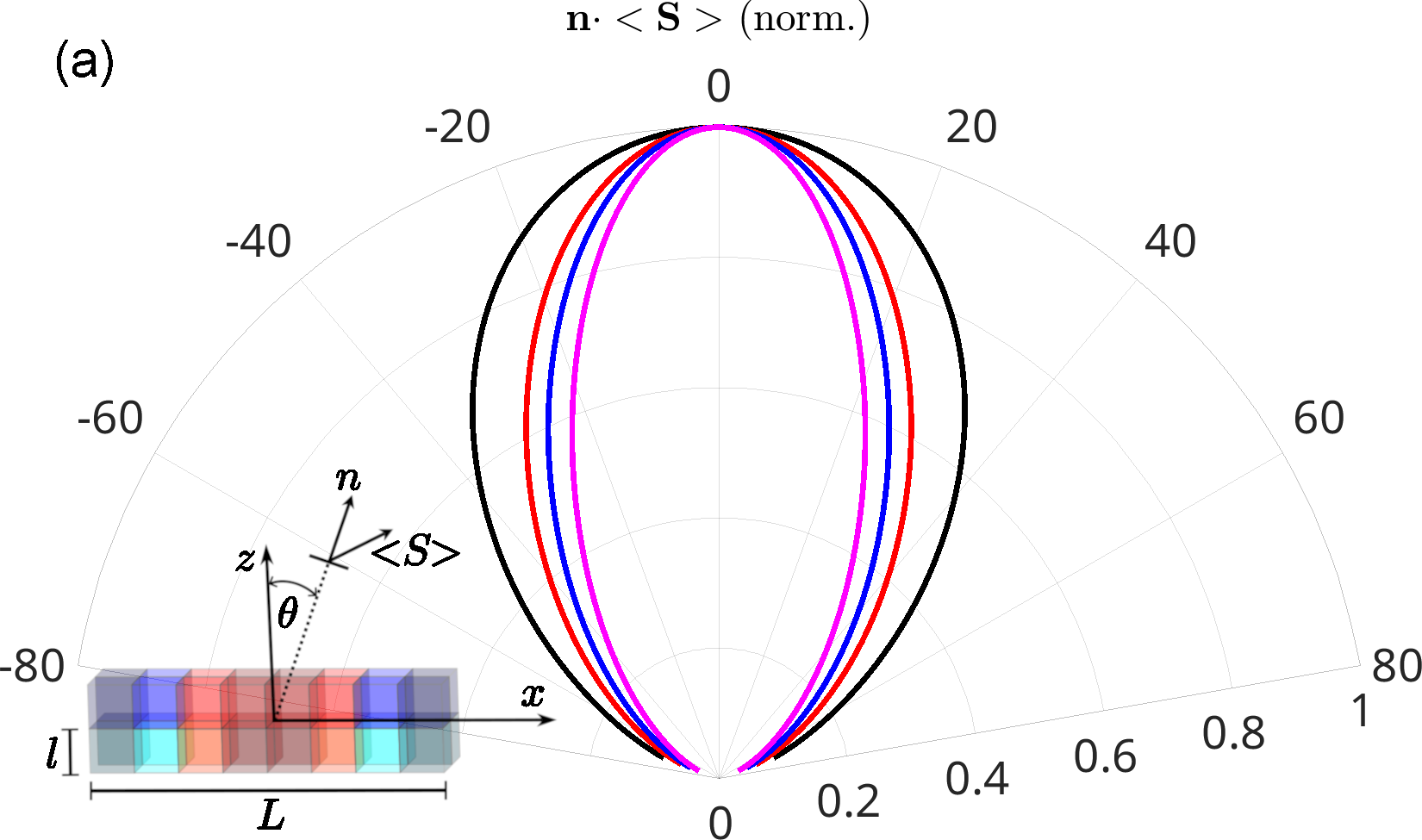}
	\includegraphics[width=0.45\textwidth]{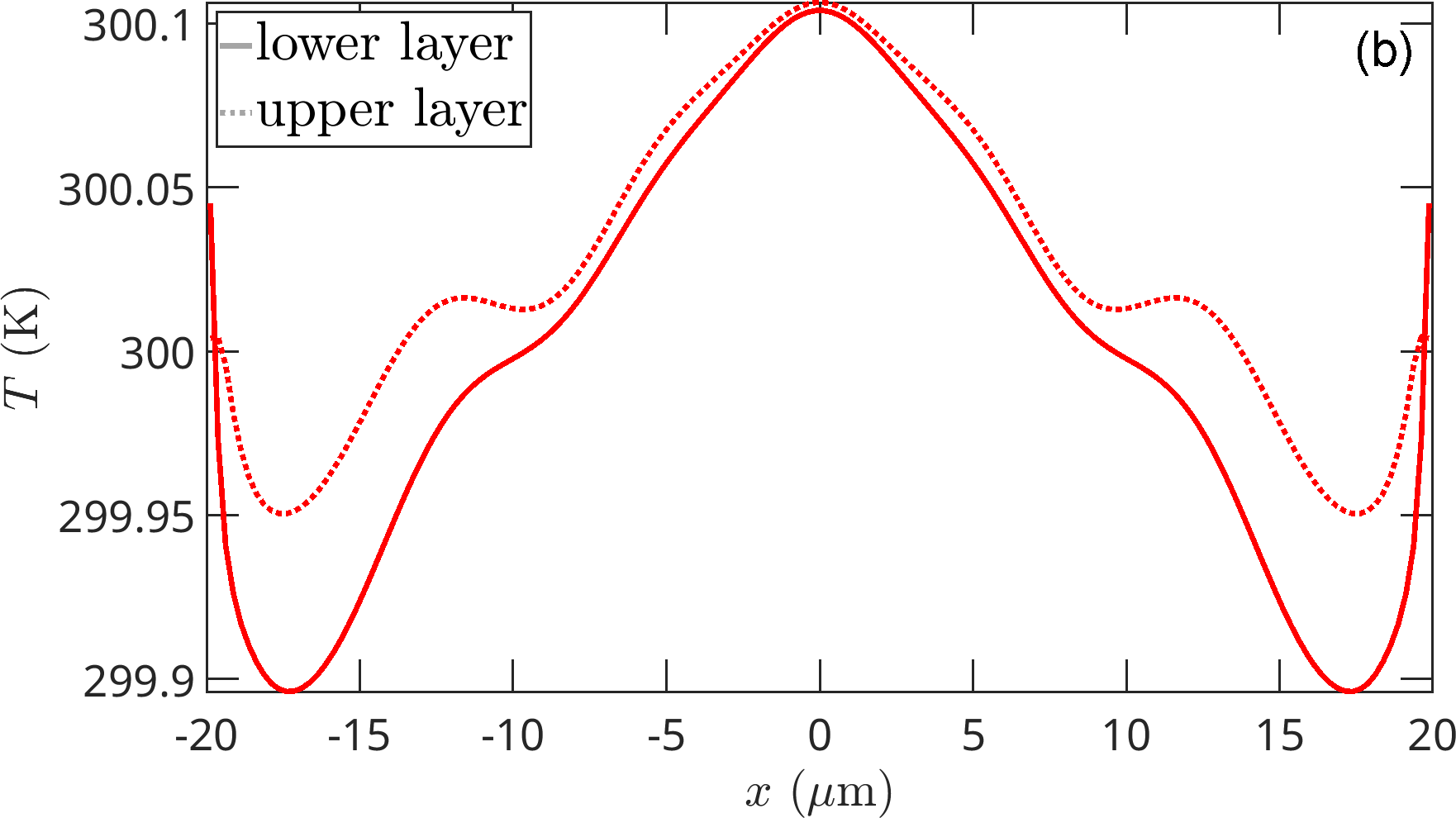}
	\includegraphics[width=0.45\textwidth]{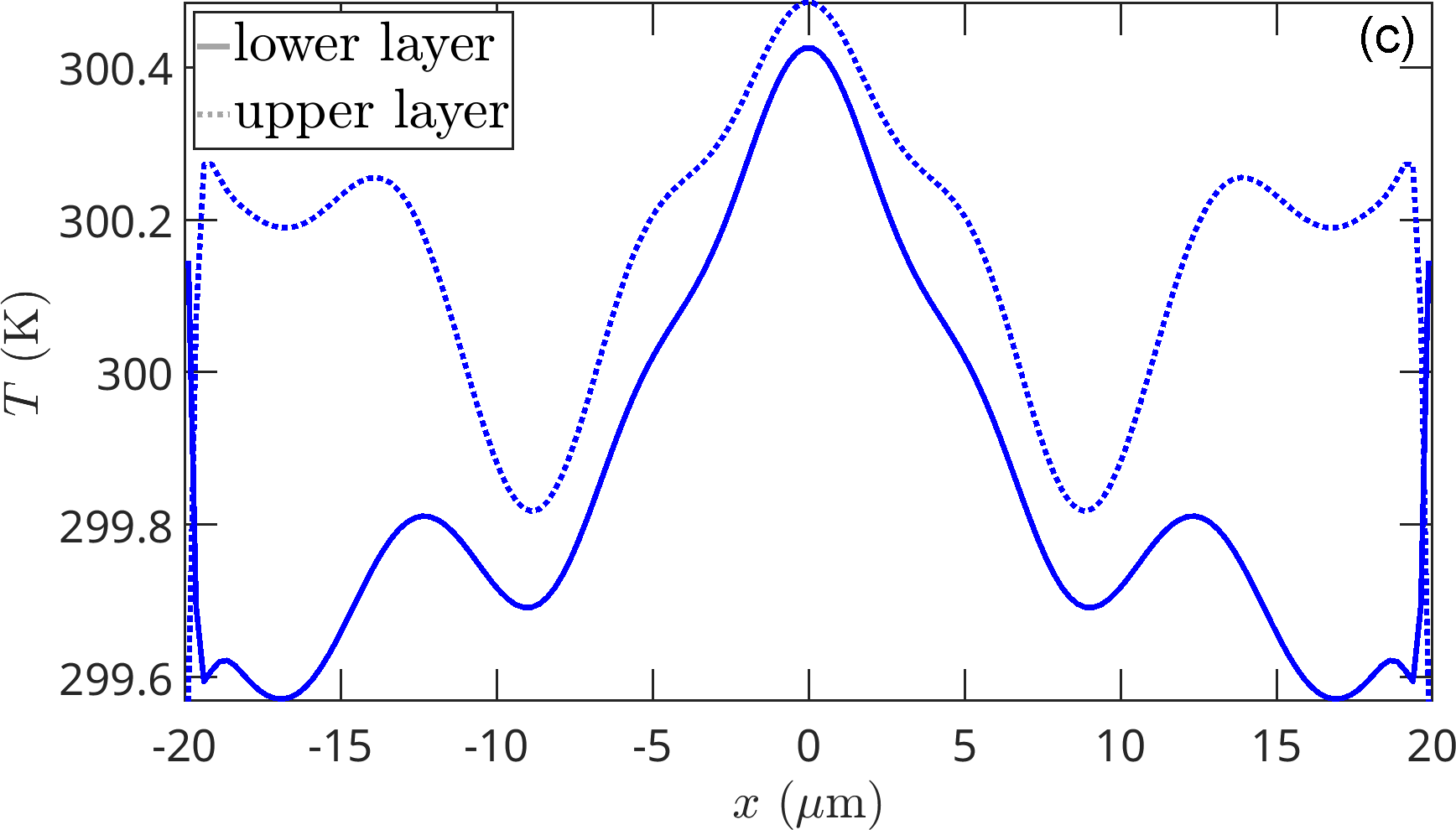}
	\includegraphics[width=0.45\textwidth]{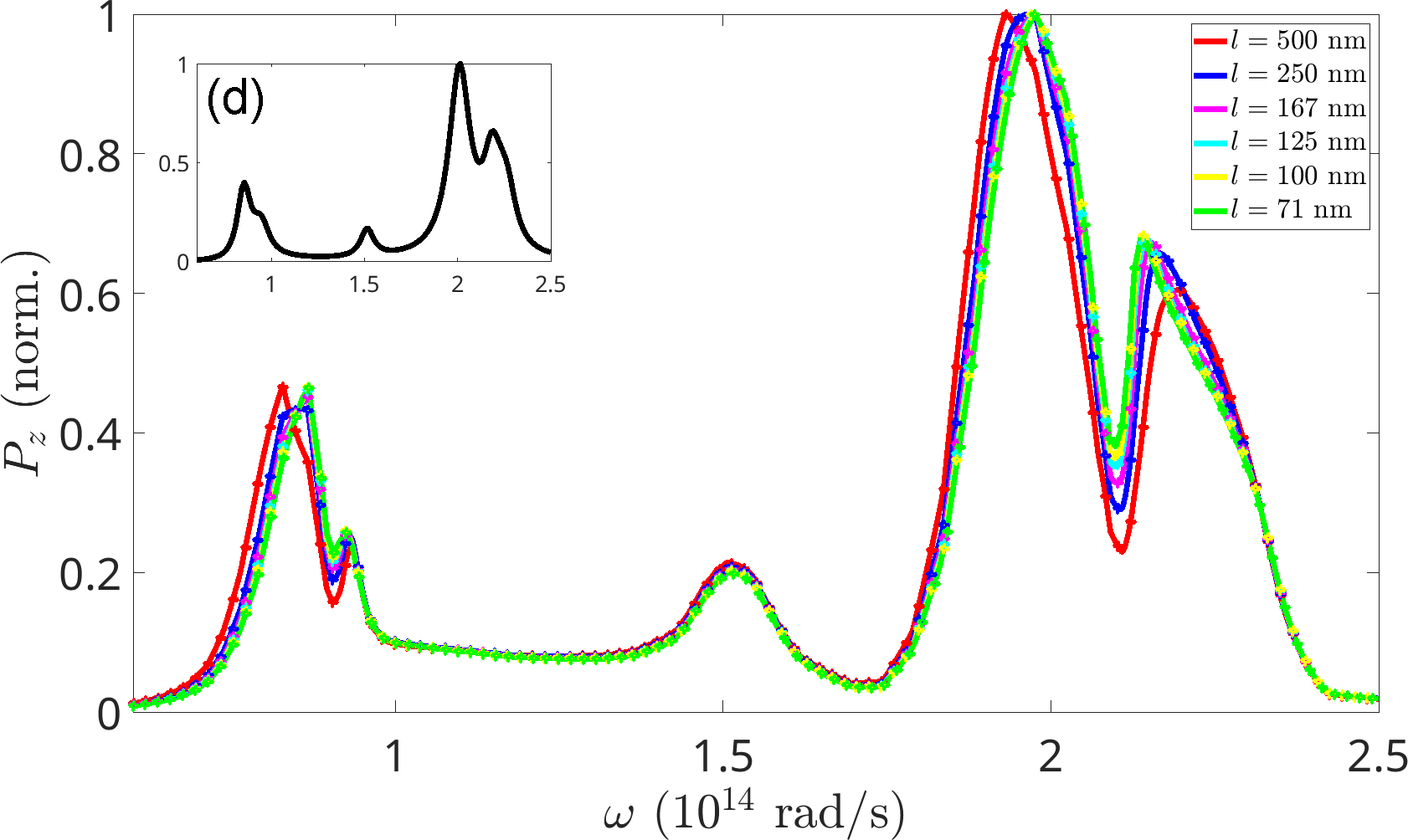}
	\caption{(a) Polar plot of Poynting flux (target) through an oriented surface of normal $\mathbf{n}=(\sin\theta,\cos\theta)$ at $z=50$ \textmu m above a SiO$_2$ parallelepiped of length $L=40$ \textmu m and of square section of side $2l=500$ nm discretized into $N=640$ cubic dipoles. Inset: sketch of emitter and its discretization into elementary cubic dipoles. (b), (c): Temperature profiles calculated along the parallelepiped for the upper and lower dipolar layers to achieve the targeted thermal emission shown in panel (a) when the bath temperature is at $T_\mathrm{b}=300$ K. (d) Spectrum of the normal component $P_z$ of Poynting vector above the source calculated from the DDA. Inset: spectrum calculated with the SCUFF-EM numerical framework~\cite{Rodriguez}.}\label{Fig1}\end{figure}
		
The heat flux radiated by the solid through an oriented surface can be calculated from the statistical averaging of the spectral Poynting vector
\begin{equation}
 \langle\mathbf{S}(\mathbf{r},\omega)\rangle=2\, \mathrm{Re}\langle\mathbf{E}(\mathbf{r},\omega)\times \mathbf{H^*}(\mathbf{r},\omega)\rangle.
\label{Poynting}
\end{equation}
From Eqs.~\eqref{Eq:field_fluc} and \eqref{Eq:mag_fluc} and using the fluctuation dissipation theorem~\cite{messina2013,Callen}
\begin{equation}\begin{split}
&\langle p^\mathrm{(fl)}_{i,\alpha}(\omega)p^\mathrm{(fl)*}_{j,\beta}(\omega')\rangle=2\hbar\epsilon_0\chi_i(\omega)\Theta(T_{i},\omega)\delta_{ij}\delta_{\alpha\beta}2\pi\delta(\omega-\omega'),\\
&\langle E_{\mathrm{b},\alpha}(\mathbf{r},\omega) H^*_{\mathrm{b},\beta}(\mathbf{r}',\omega')\rangle=2i\hbar\mu_0\omega^2\Theta(T_\mathrm{b},\omega)\\
&\qquad\times\mathrm{Re}[\mathds{G}^{\mathrm{(HE)}}_{\beta \alpha}(\mathbf{r},\mathbf{r}',\omega)]2\pi\delta(\omega-\omega'),
\label{FDT}
\end{split}\end{equation}
where $\chi_i = \mathrm{Im}(\alpha_i) - |\alpha_i|^2\omega^3/6\pi c^3$, $\alpha_{i}$ being the polarizability associated to the $i$-th dipole, and $\Theta(T,\omega)={\hbar\omega}/[\exp(\hbar\omega/k_B T)-1]$ is the mean energy of a harmonic oscillator at temperature $T$, it is straightforward to show that each component of this vector can be recast into
\begin{equation}
\langle S_{\zeta}(\mathbf{r},\omega)\rangle= \sum_{i=1}^Na_{\zeta}(\mathbf{r},\mathbf{r}_i,\omega)\bigl[\Theta(T(\mathbf{r}_i),\omega)-\Theta(T_\mathrm{b},\omega)\bigr],
\label{Poynting_bis}
\end{equation}
where
\begin{equation}\begin{split}
a_{\zeta}(\mathbf{r},\mathbf{r_i},\omega)&=\epsilon_{\zeta\gamma\beta}\frac{4\hbar\omega^4}{\epsilon_0c^4}\chi_i(\omega)\\
&\,\times\mathrm{Re}[\mathds{G}^{(\mathrm{EE})}_{\gamma\eta}(\mathbf{r},\mathbf{r}_i,\omega)\mathds{G}^{\mathrm{(HE)}*}_{\beta\eta}(\mathbf{r},\mathbf{r}_i,\omega)],
\label{kernel}
\end{split}\end{equation}
in which a summation is intended over the repeated indices.
\subsubsection*{B. Inverse problem: retrieve the temperature profile}
\begin{figure}
	\centering
	\includegraphics[width=0.45\textwidth]{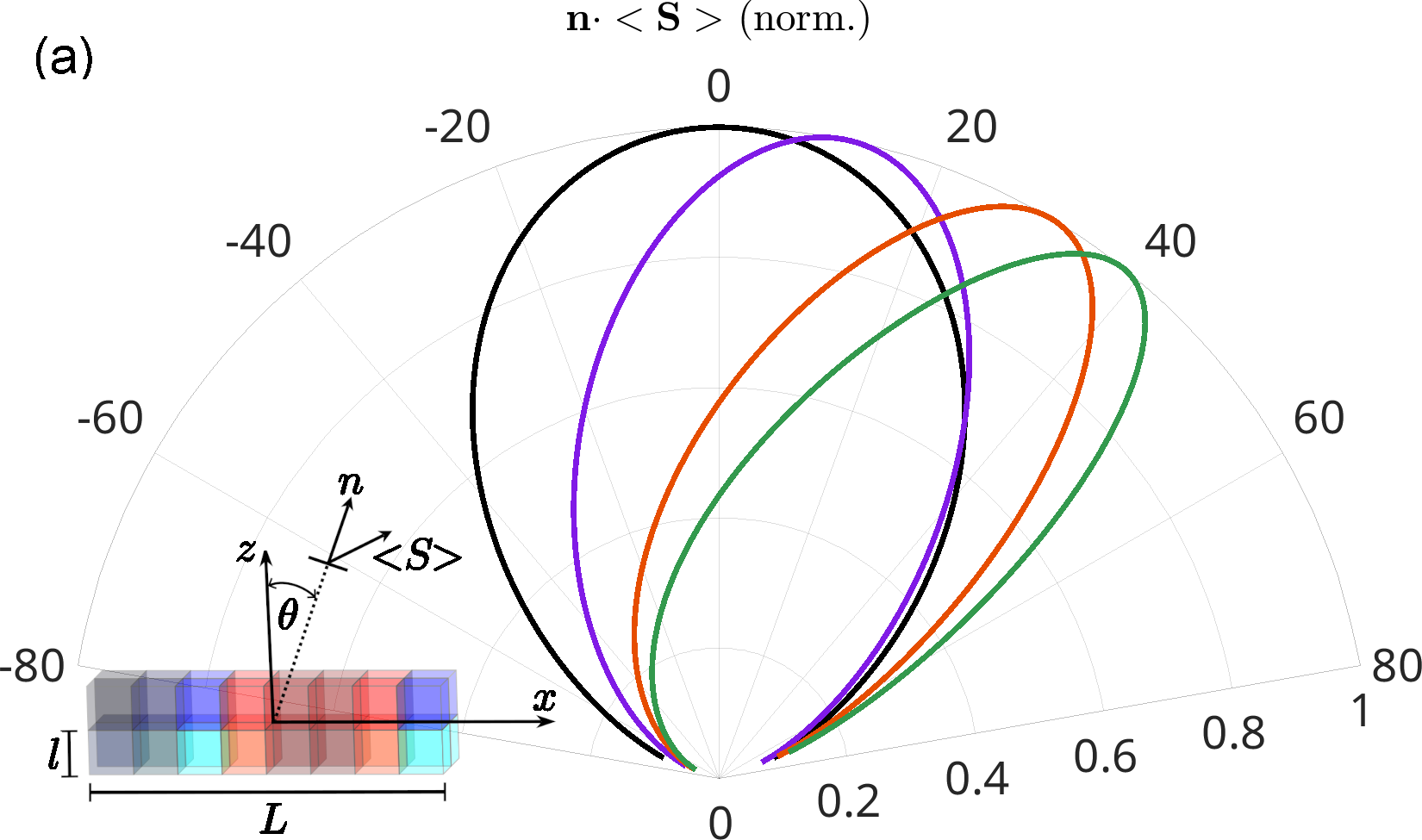}
	\includegraphics[width=0.45\textwidth]{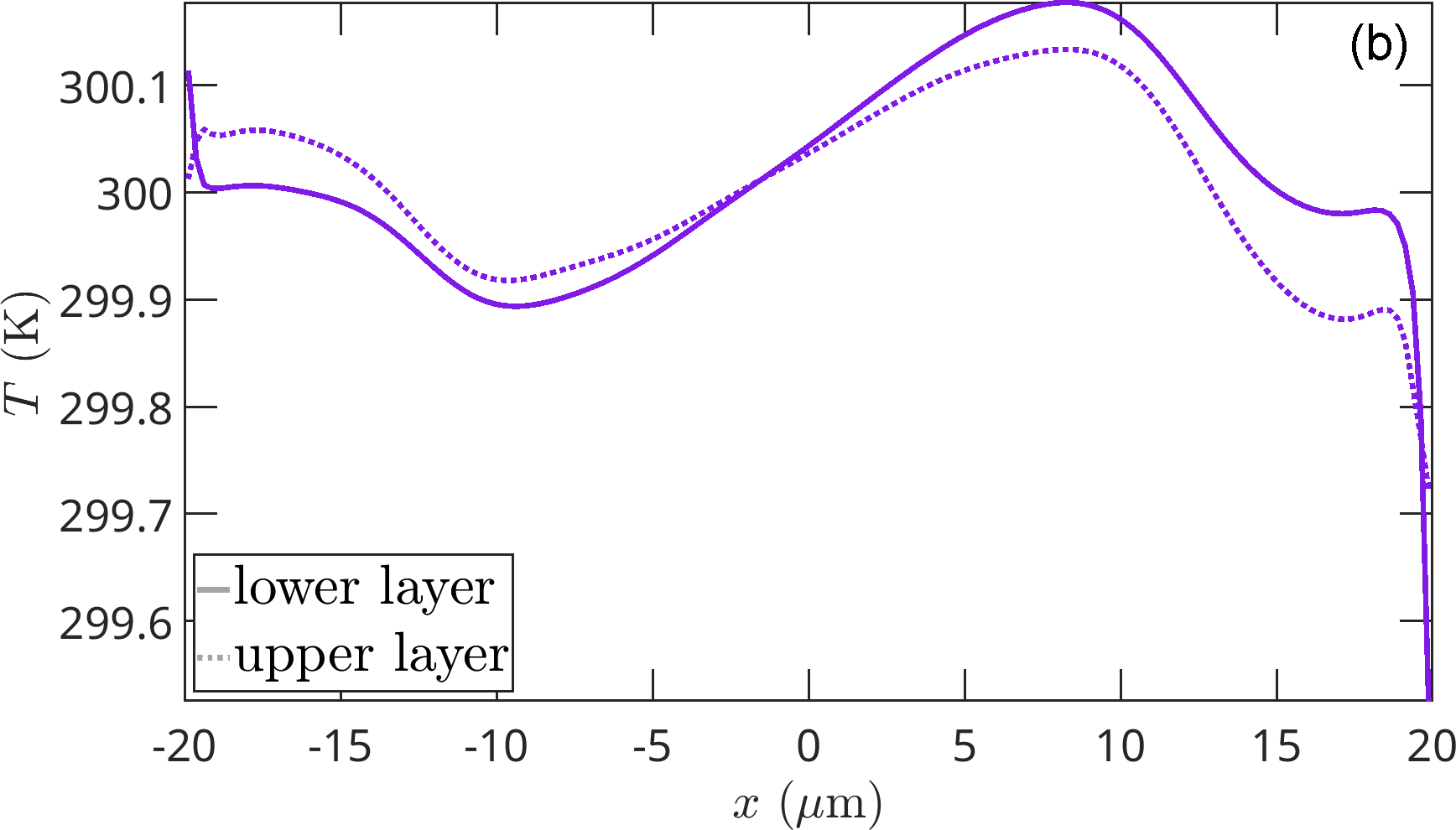}
	\includegraphics[width=0.45\textwidth]{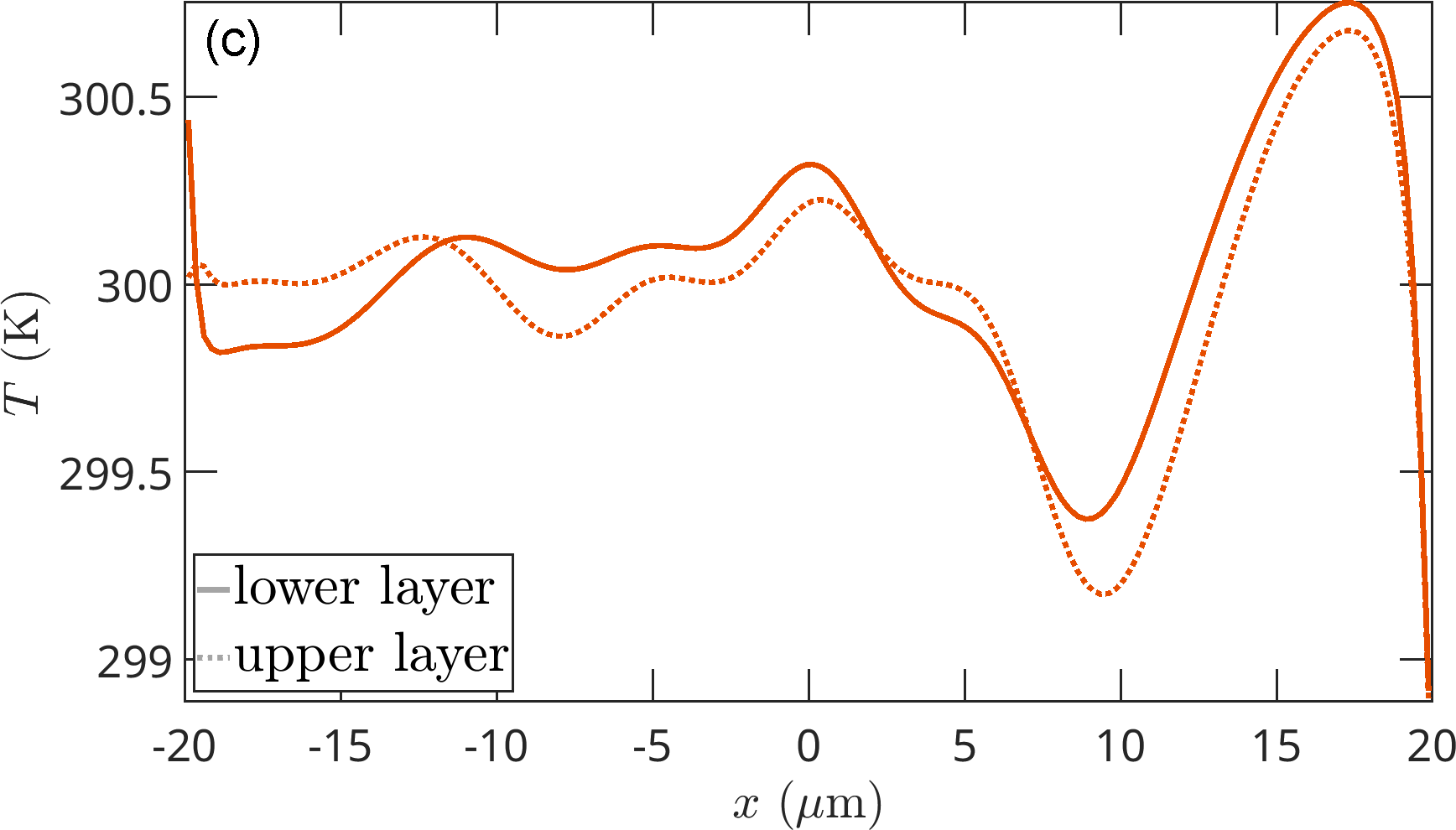}
	\caption{(a) Polar plot of a tilted Poynting flux (target) through an oriented surface of normal $\mathbf{n}=(\sin\theta,\cos\theta)$ at $z=50\,\mu$m above a SiO$_2$ parallelepiped of length $z=40\,\mu$m and of square section of side $2l=500$ nm discretized into $N=640$ cubic dipoles. Inset: sketch of emitter and its discretization into elementary cubic dipoles. (b), (c): Temperature profiles calculated along the parallelepiped for the upper and lower dipolar layers to achieve the tilted thermal emission emission shown in panel (a) when the bath temperature is $T_\mathrm{b}=300\,$K.}\label{Fig2}
\end{figure}

 Equation~\eqref{Poynting_bis} established a close relation between the temperature profile within the solid and the spectral components of the Poynting vector radiated in its surrounding. It shows that the multi-scattering process described by the full Green tensor can be weighted by the temperature of local thermal emitters within a solid. Assuming a weak temperature variation within the system close to the environmental temperature $T_\mathrm{b}$, the full Poynting vector $\langle\mathbf{S}(\mathbf{r})\rangle=\int\frac{d\omega}{2\pi}\langle\mathbf{S}(\mathbf{r},\omega)\rangle$ can be recast in the following form
\begin{equation}
\langle\mathbf{S}(\mathbf{r})\rangle=\sum_{i=1}^N\int\frac{d\omega}{2\pi}\, \mathbf{a}(\mathbf{r},\mathbf{r}_i,\omega)\frac{d\Theta(T_\mathrm{b},\omega)}{dT}[T(\mathbf{r_i})-T_\mathrm{b}],
\label{linear}
\end{equation}
which linearly relates the Poynting vector radiated by the system to its temperature profile. This expression is the main result of this work and it can be used in inverse design to determine the precise temperature profile required to achieve a desired emission pattern. In Figs.~\ref{Fig1} and \ref{Fig2} we illustrate the potential of inverse design to control the heat flux radiated by a simple dipolar chain made of silicon carbide (SiO$_2$) nanoparticles, whose optical properties are described by a Drude-Lorentz model~\cite{Palik}. In these examples, our objective (target) is the following Lorentzian heat flux crossing the oriented surfaces of normal unit vector $\mathbf{n}$ at a distance $z$ from the structure 
\begin{equation}
\mathbf{n}\cdot\langle\mathbf{S}_\mathrm{targ}\rangle=\frac{S_0}{1+(\frac{x-x_0}{\Gamma})^2}.
\label{Lorentzian}
\end{equation}
Here $\Gamma$ and $S_0$ denote the half-width at half-maximum (HWHM) of the Lorentzian and the maximum flux while $x_0$ sets the direction $\theta_0$ of emitted flux ($x_0=z\tan \theta_0$). The inverse problem is solved using a Tikhonov's regularization method~\cite{Tikhonov} to identify the temperature profiles necessary to reach a Lorentzian flux. Tikhonov's regularization is used to stabilize the solution of ill-posed inverse problems, where small errors in input data can lead to large errors in the output due to badly conditioned matrices. The main idea is to modify the original problem [i.e. solving Eq.~\eqref{linear} with respect to the temperature profile for a given target \eqref{Lorentzian}] by adding a penalty term that discourages large fluctuations in the solution. Specifically, instead of minimizing the difference between the calculated emission profile and the targeted emission, the Tikhonov regularization minimizes a combination of this difference and a regularization term that imposes smoothness. Mathematically, this is typically written as minimizing the functional $F(\mathbf{T})=\left\lVert \mathbf{A}\cdot\mathbf{T}- \mathbf{n}\cdot\langle\mathbf{S}_\mathrm{targ}\rangle \right\rVert^2+\alpha \left\lVert \mathbf{T} \right\rVert^2$, where $\mathbf{A}$ is the matrix coming from the discretization of Eq. (7), $\mathbf{T}$ the temperature vector providing the solution of the problem, and $\alpha$ is a regularization parameter that balances data fidelity and regularization. By choosing an appropriate $\alpha$ (see Ref.~\onlinecite{Tikhonov}), the Tikhonov method produces stable solutions as a minimum of the convex functional $F$.

\begin{figure}
	\centering
	\includegraphics[width=0.45\textwidth]{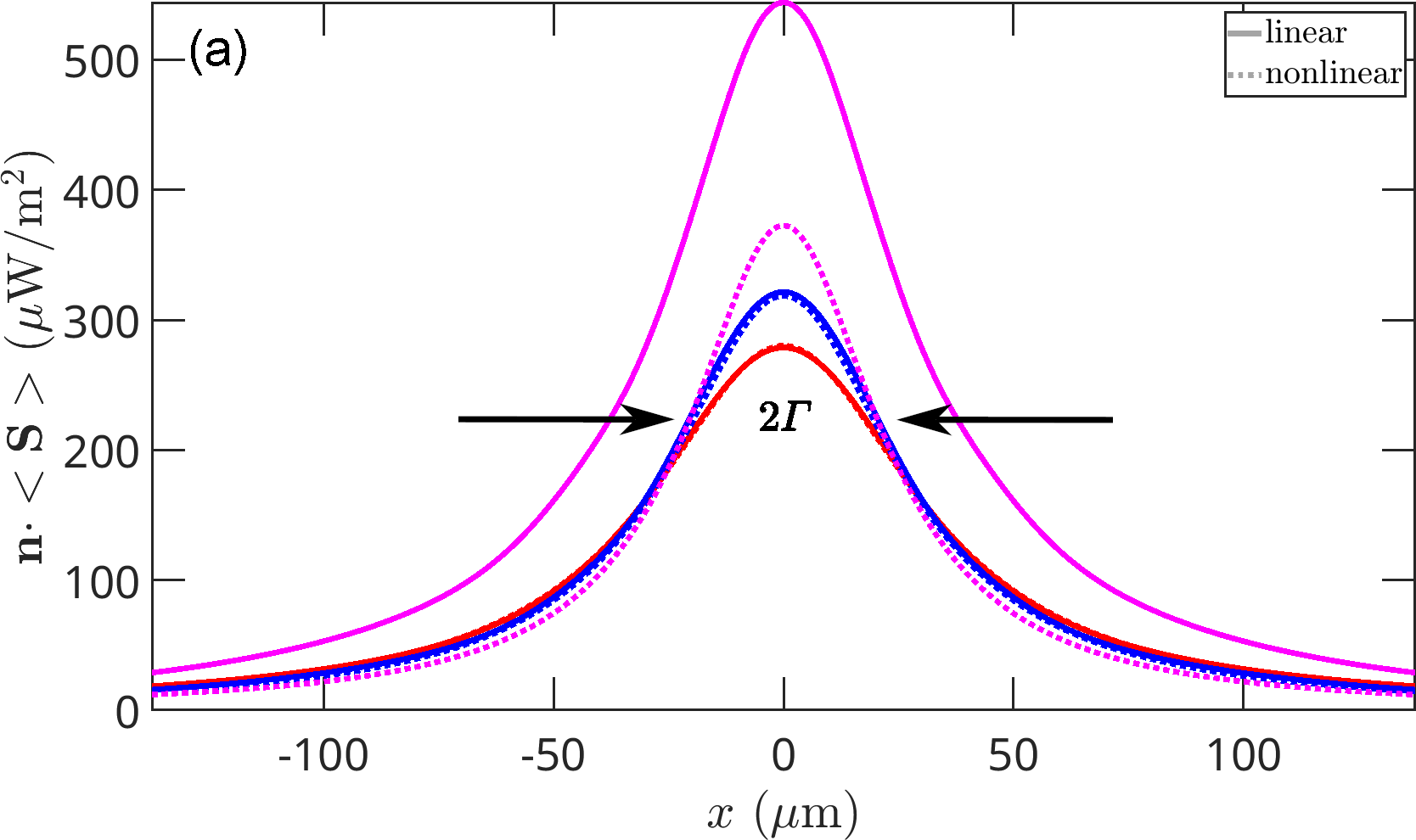}
	\includegraphics[width=0.45\textwidth]{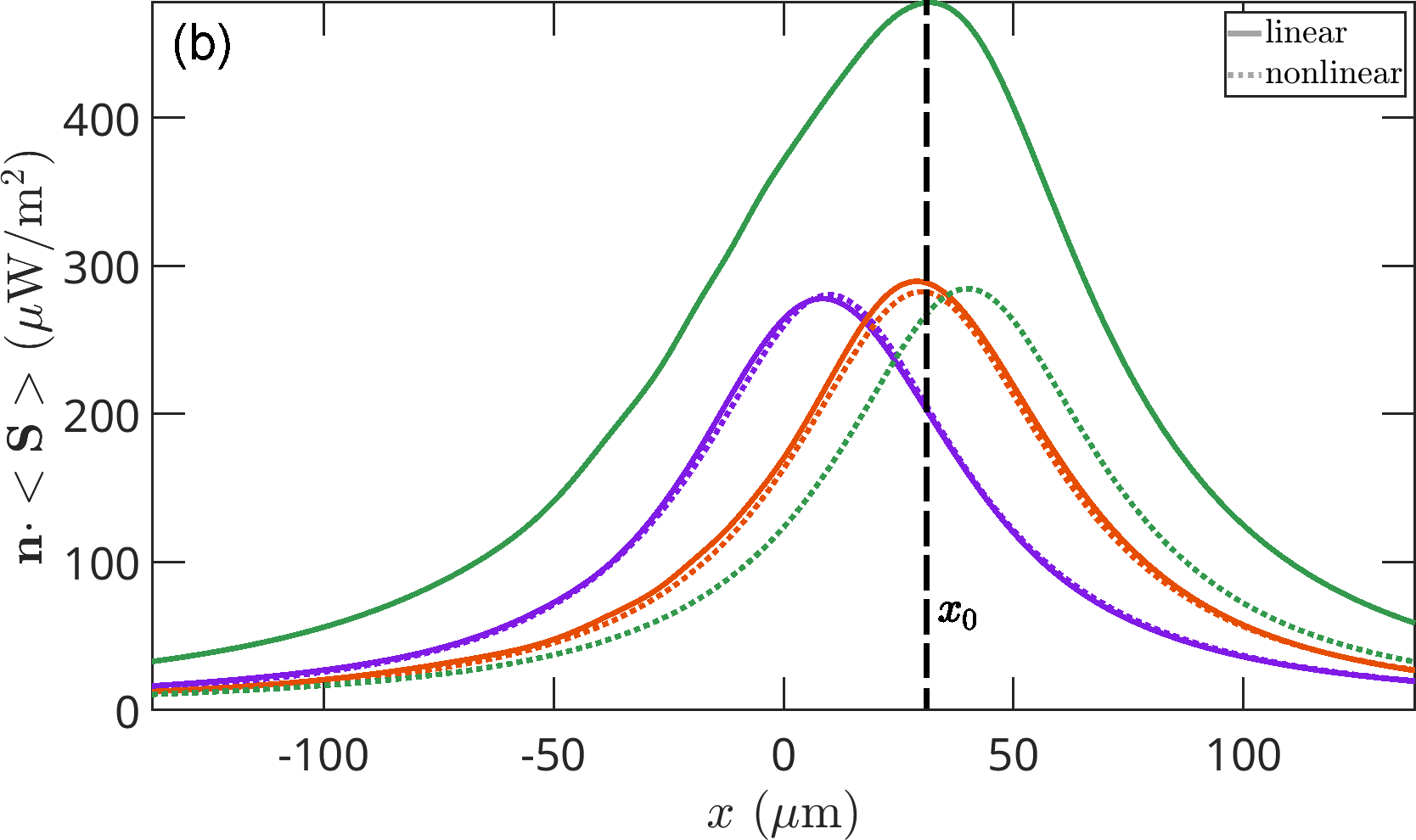}
	\caption{Comparison of the Poynting flux radiated by the same parallelepiped as in Figs.~\ref{Fig1} and \ref{Fig2} calculated in the linear approximation (dashed lines) and using the exact (nonlinear) model (solid lines) for the normal emission pattern (a) and the tilted one (b). The line colors correspond to the targets defined in Figs.~\ref{Fig1} and \ref{Fig2}.}\label{Fig3}
\end{figure}

The solution of this problem is plotted in Figs.~\ref{Fig1} in the case of normal emission (i.e. $x_0=0$) and in Figs.~\ref{Fig2} for a tilted emission, for various HWHM. The thermal emitter we consider here is a parallelepiped of length $L = 40$ \textmu m and of square section of side $2l = 500$ nm. It is discretized into $640$ elementary emitting cubic dipolar volumes regularly distributed as sketched in the insets of Figs.~\ref{Fig1}(a) and ~\ref{Fig2}(a). Notice that randomness of this local emitters does not significantly affect the final outcome but only impacts the speed of convergence of the calculation method. The different targets which correspond to the emission pattern at $z=50$ \textmu m above the parallelepiped are plotted in Figs.~\ref{Fig1}(a) and \ref{Fig2}(a) for a normal and tilted emission, respectively. As expected, the temperature profiles, solution of the inverse problem, are symmetric with respect to the center of the parallelepiped for a normal emission pattern [Fig.~\ref{Fig1}(b)] while it is no longer symmetric for a tilted emission [Fig.~\ref{Fig2}(b)]. Furthermore, we see that, to reduce the width of emission pattern and, therefore, focus heat in a narrower lobe, the temperature contrast between the center of emitter and its edges must be increased. In Fig.~\ref{Fig1}(d) we show the spectrum of the Poynting vector above the emitter in normal incidence calculated from the discrete dipolar approximation (DDA)~\cite{Pennypacker} and compare it (see inset of the same figure) with a calculation performed with SCUFF-EM, a free open-source software based on the boundary element method~\cite{Rodriguez}. This spectrum clearly demonstrates the broadband nature of the emission. Notice also that, when the local temperature variation within the system becomes too significant, that is when the temperature gradient is large enough, the solution to the problem is no longer compatible with the linear approach we have adopted to solve the inverse problem. Indeed, as shown in Fig.~\ref{Fig3} we obtain a discrepancy between the emission patterns calculated in the linear approximation of the Poynting vector with Eq.~\eqref{linear} and that one obtained from the exact expression Eq.~\eqref{Poynting_bis}. These curves confirm that the linear approach is sufficient if the target emission can be obtained by weakly varying temperature profiles. This sets fundamental limits on the control of the heat flux by anisothermal sources (at least within the linear approximation). Consequently, in this study, we intentionally confine our analysis to the linear regime in order to simplify the resolution of inverse problem—namely, the inversion of Eq.~\eqref{linear}. Without this approximation, the problem becomes nonlinear and is considerably more challenging to solve. Although a more general nonlinear framework could, in principle, be used to take into account the presence of arbitrary large temperature gradients, such an extension presents a primarily mathematical challenge rather than a physical interest. Indeed, at the moment, the resolution of nonlinear inverse problems of this kind remain largely open in the mathematical literature. Moreover, on a more physical point of view, the presence of conduction within the solid naturally smooths the temperature profiles and tends to reduce strong temperature variations.

\begin{figure}
\centering
\includegraphics[width=0.45\textwidth]{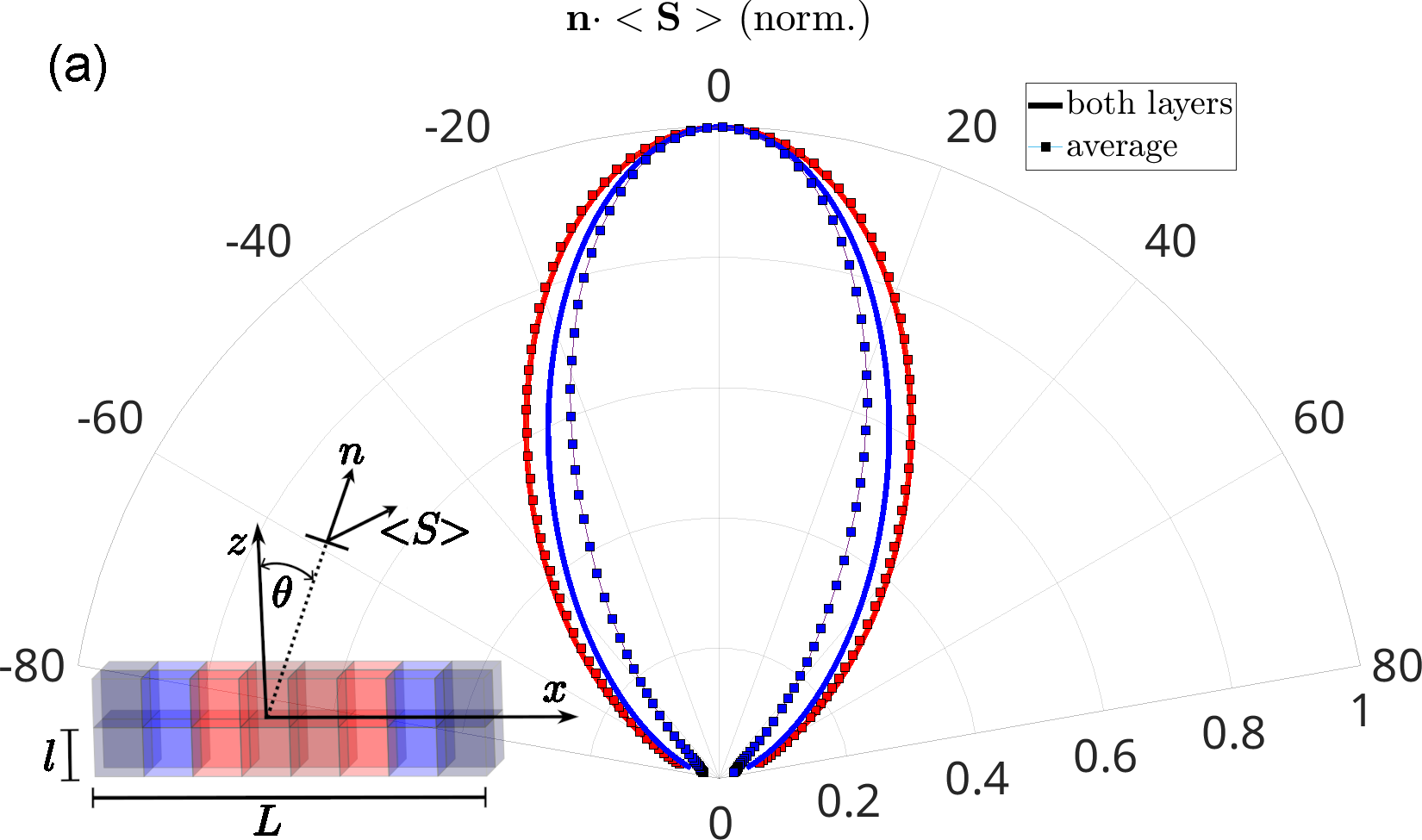}
\includegraphics[width=0.45\textwidth]{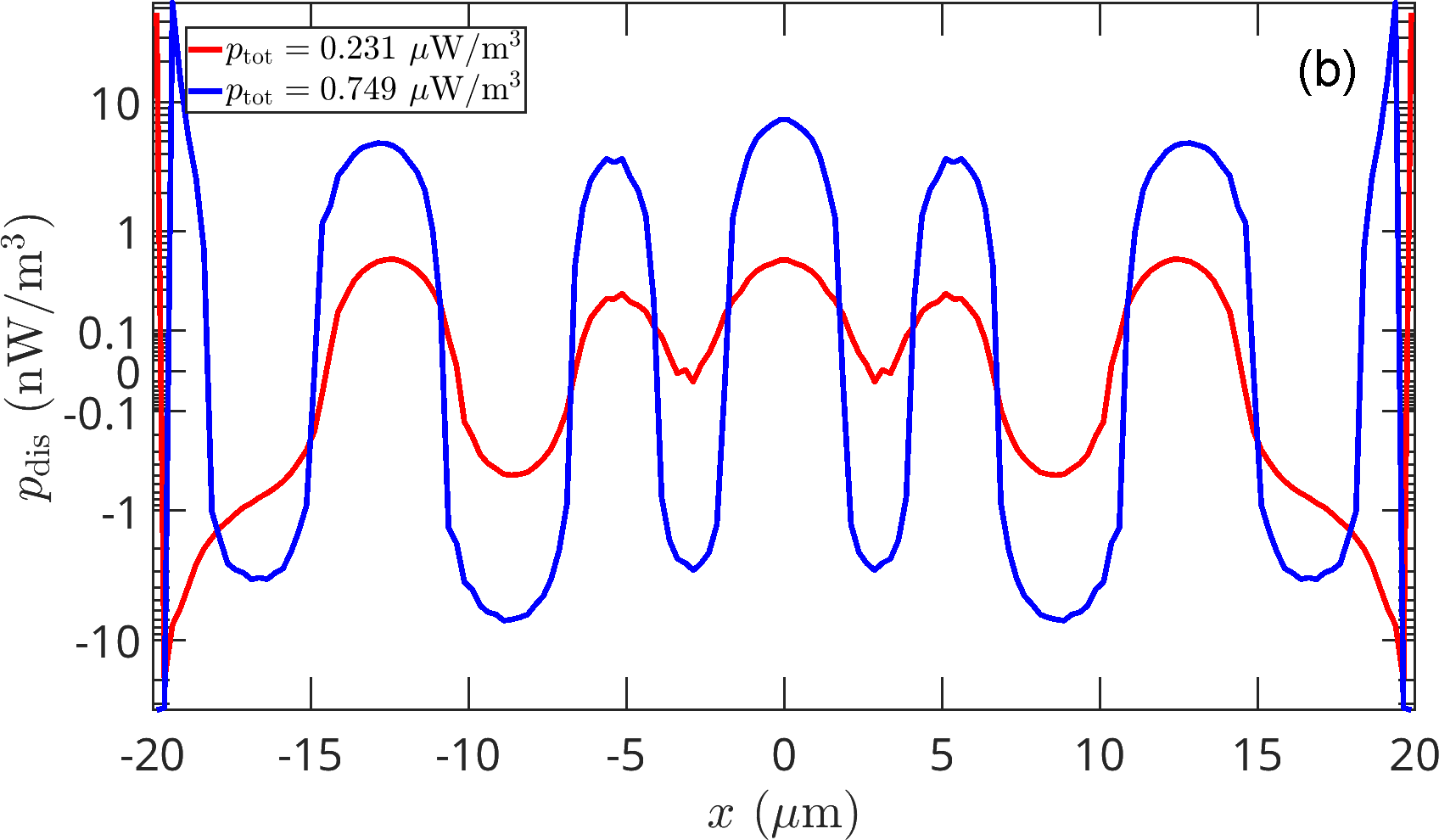}
\caption{Power density required along the same SiO$_2$ parallelepiped as in Fig.~\ref{Fig1} to achieve the target emission. (a) Target emission patterns. The solid line corresponds to the target leading to the temperatures profiles plotted in Fig~\ref{Fig1}(b). The dashed curve corresponds to the emission pattern obtained by taking the average of these two temperatures profiles. (b) Spatial distribution of the power density along the parallelepiped on the external face of the lower layer to achieve the dashed targeted emission. The thermal conductivity of SiO$_2$ is $\lambda=1.3\,\mathrm{W}\cdot\mathrm{m}^{-1}\cdot\mathrm{K}^{-1}$ (see Ref.~\onlinecite{conductivity}). Inset: total power necessary to achieve the emission control.}
\label{Fig4}
\end{figure}

\begin{figure}
\centering
\includegraphics[angle=0,scale=0.18,angle=0]{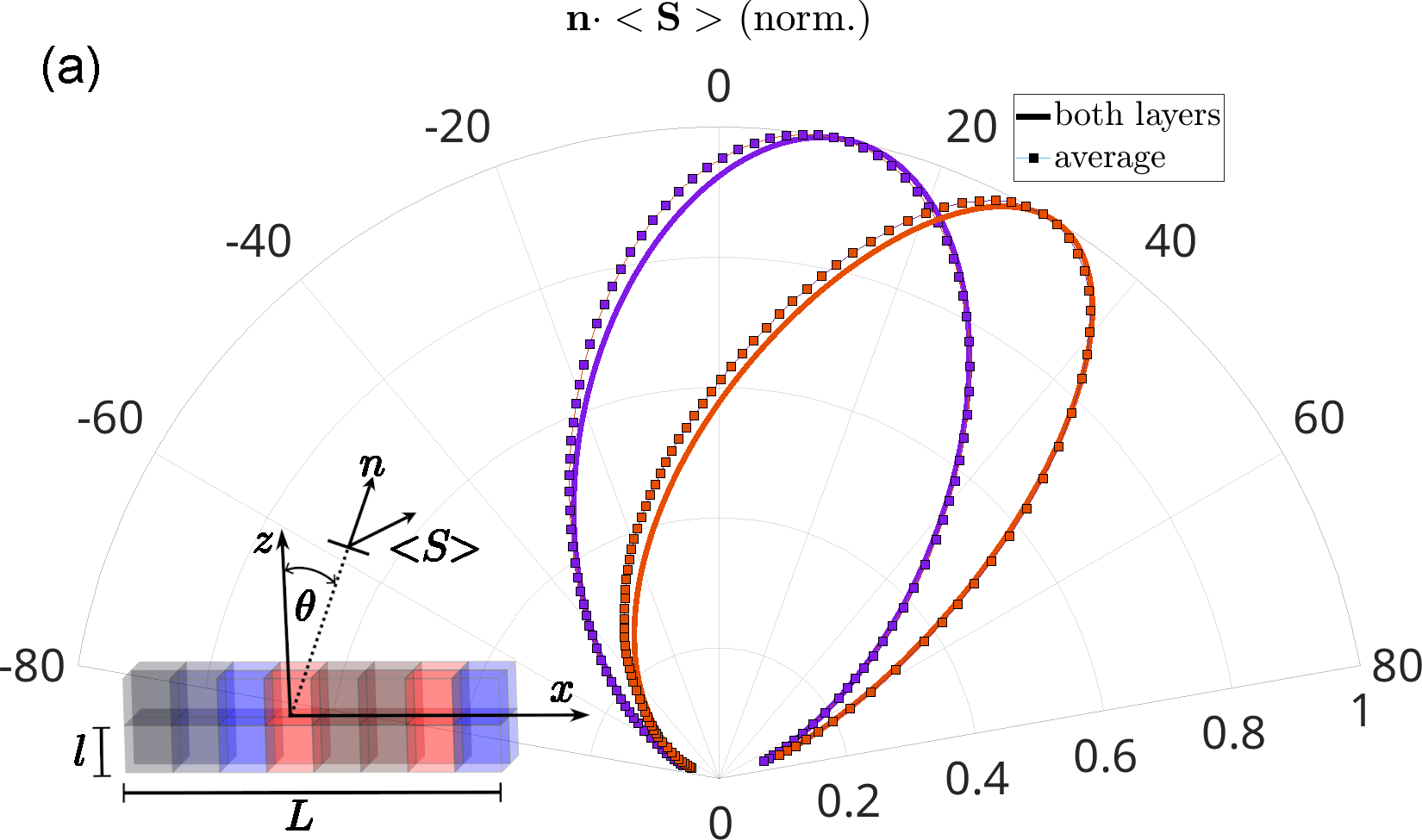}
\includegraphics[angle=0,scale=0.18,angle=0]{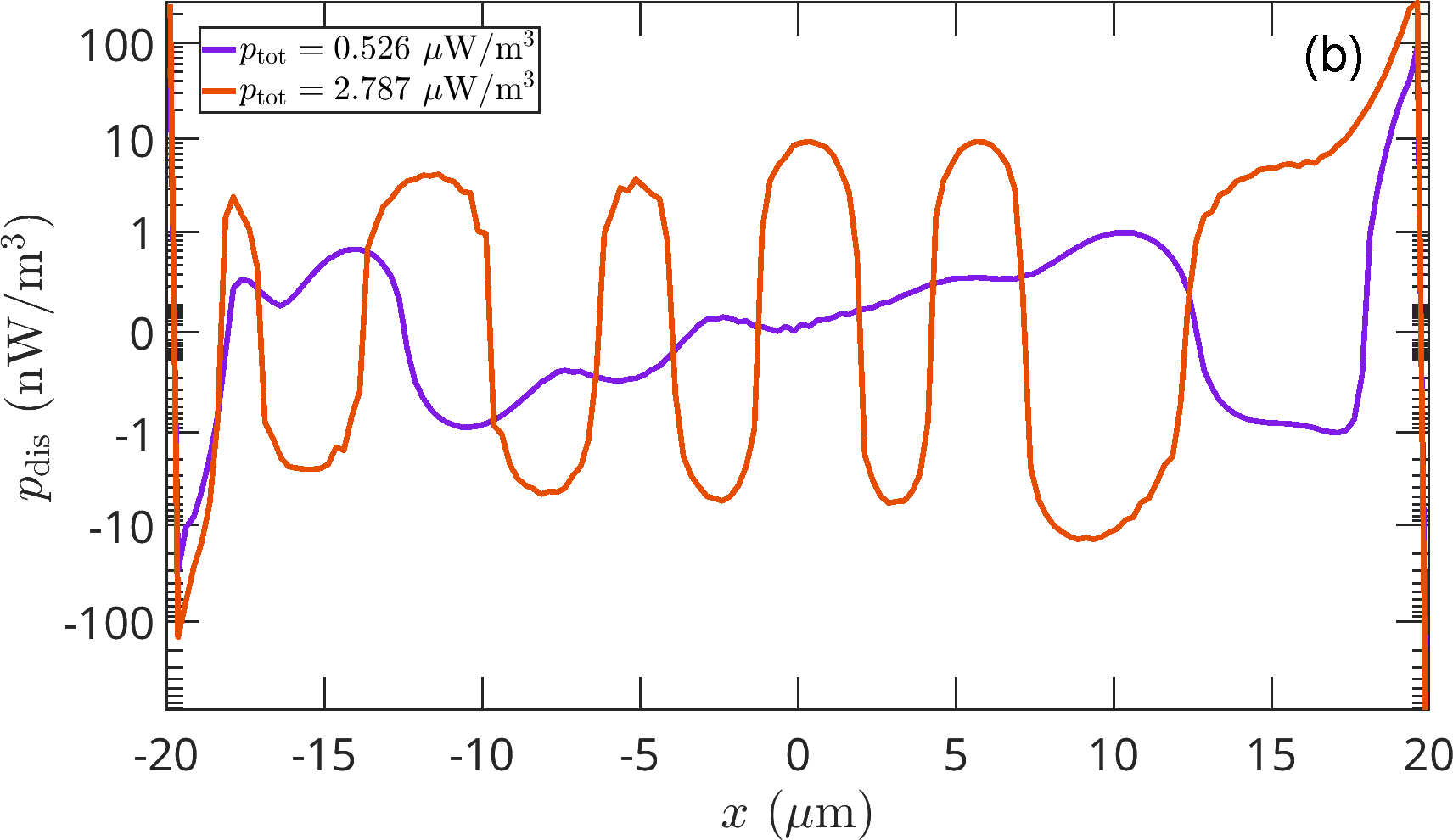}
\caption{Power density required along the same SiO$_2$ parallelepiped as in Fig.~\ref{Fig2} to achieve a tilted target emission. (a) Target emission patterns. The solid line corresponds to the target leading to the temperatures profiles plotted in Fig~\ref{Fig2}(b). The dashed curve corresponds to the emission pattern obtained by taking the average of these two temperatures profiles. (b) Spatial distribution of the power density along the parallelepiped on the external face of the lower layer to achieve the dashed targeted emission. Inset: total power necessary to achieve the emission control.}\label{Fig5}
\end{figure}
\subsection*{III. Control of temperature profile}
To reach the target emission, the temperature profile within the solid must be carefully controlled, requiring localized heating or cooling, as shown in Figs.~\ref{Fig4} and \ref{Fig5}. When the thermal source is sufficiently thin the transverse temperature gradient is small so that the temperature profile varies only laterally.

\begin{figure}[htbp]
 \centering
 \begin{tabular}{cc}
 \includegraphics[width=0.32\textwidth]{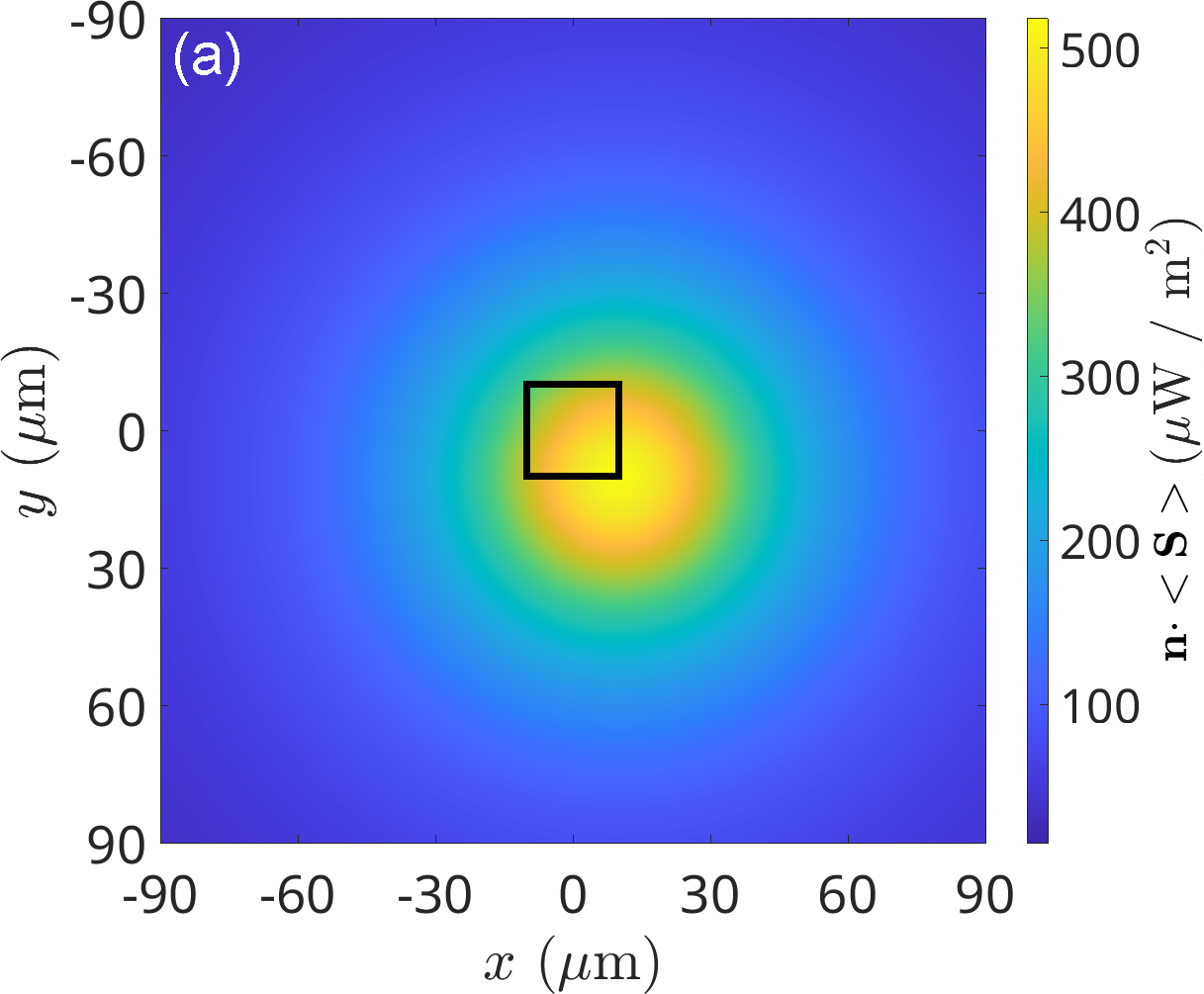}\\
 \includegraphics[width=0.34\textwidth]{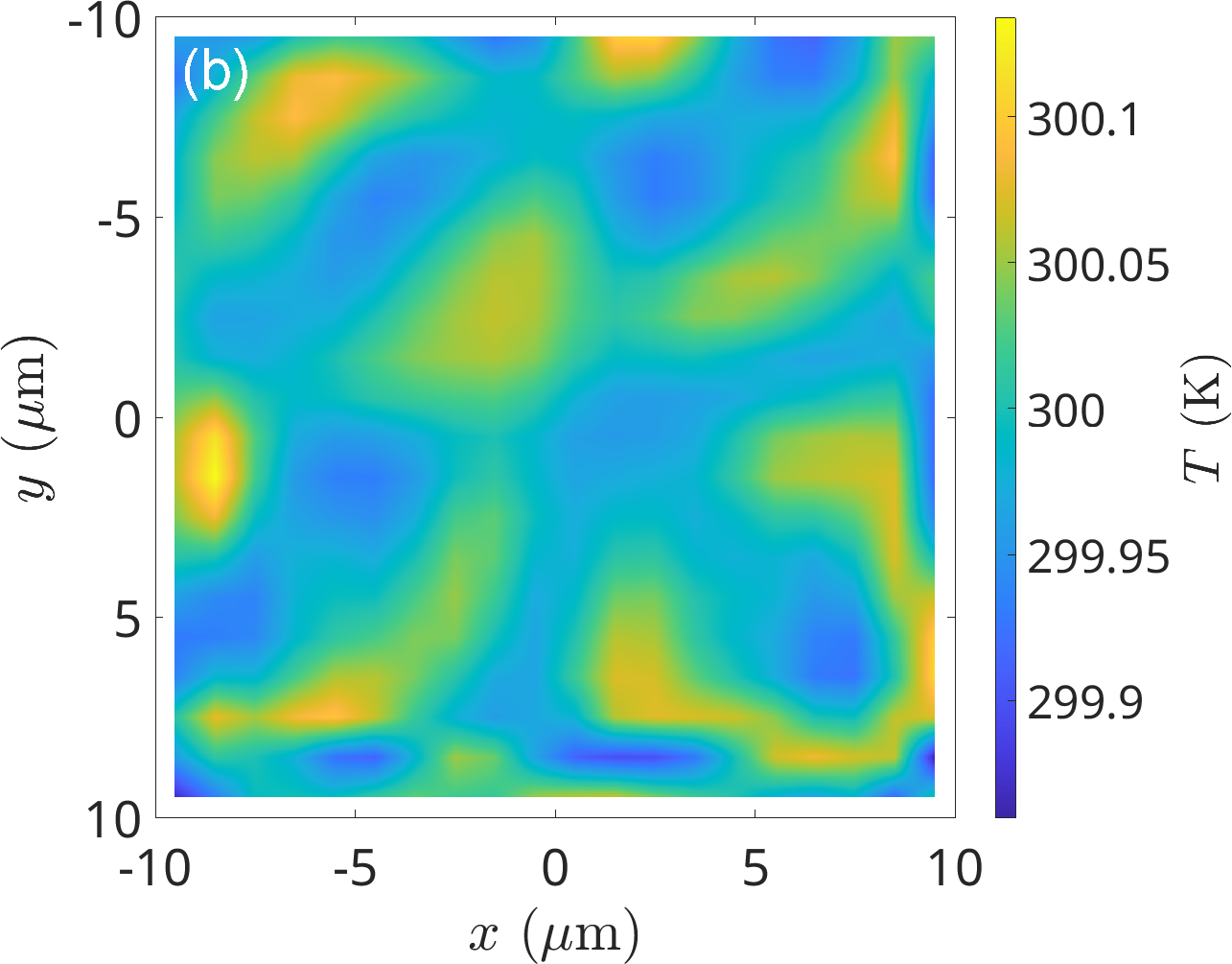}\\
 \includegraphics[width=0.32\textwidth]{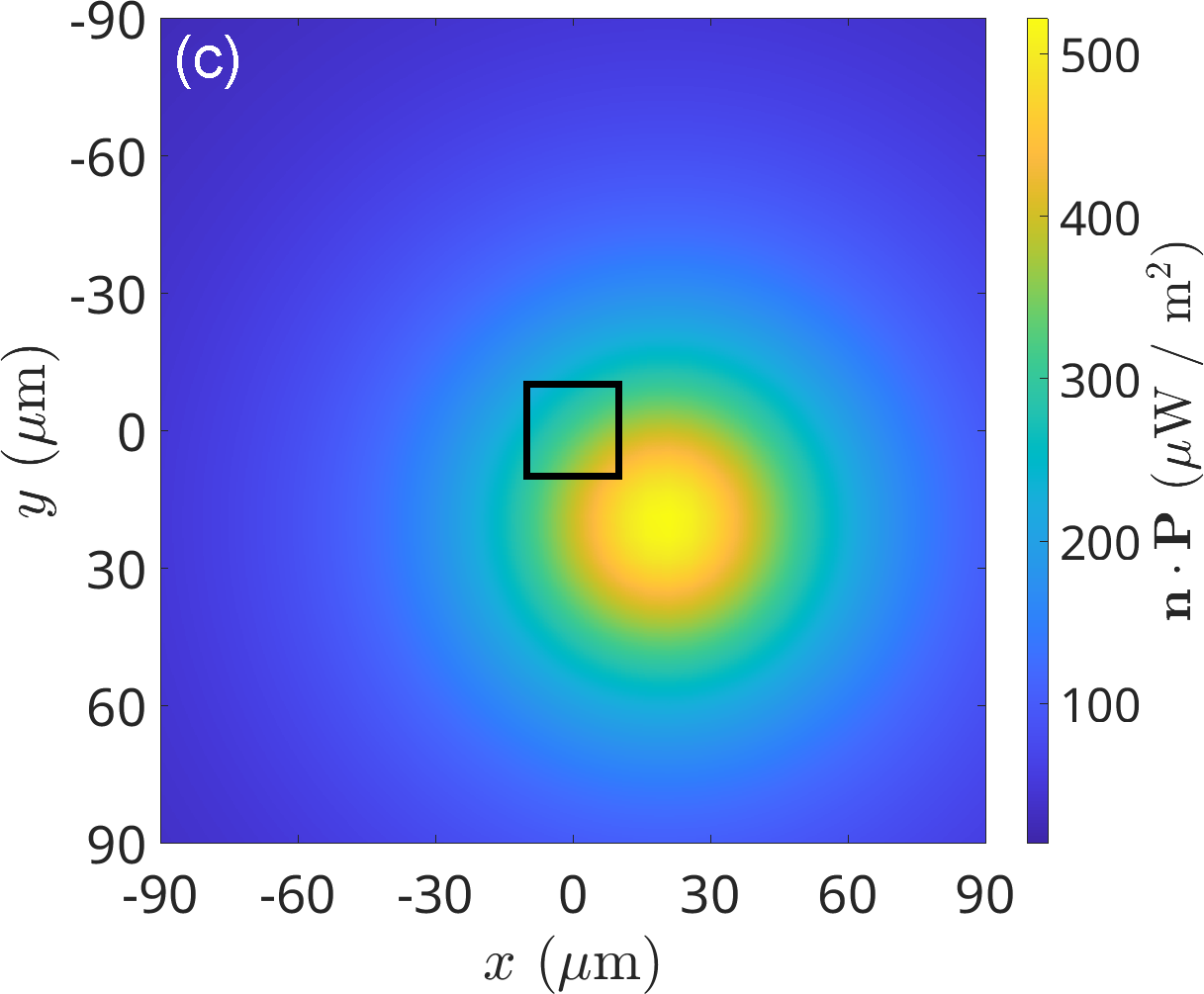} \\
 \includegraphics[width=0.34\textwidth]{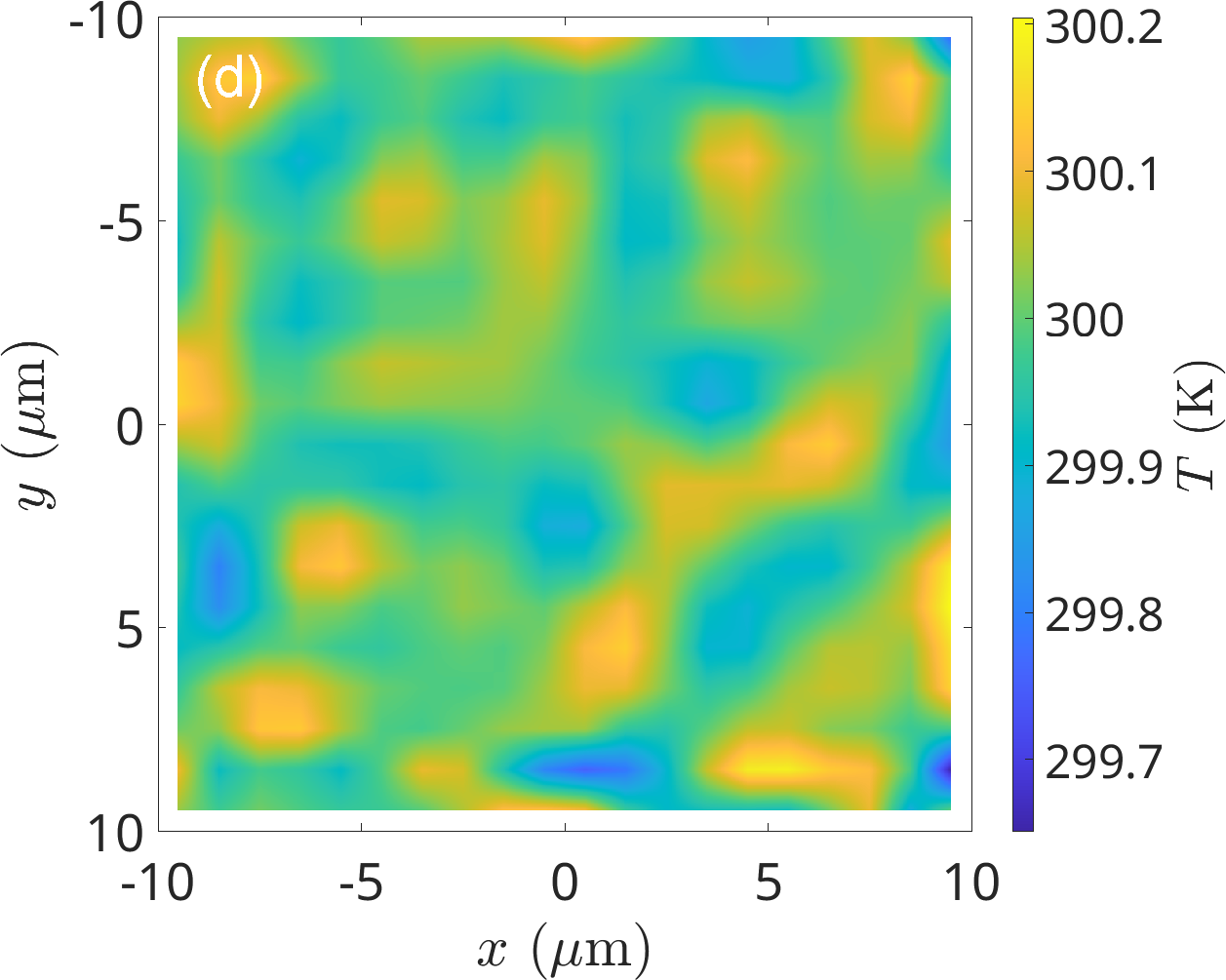}
 \end{tabular}
 \caption{Tilted emission pattern. (a) Poynting flux through an oriented surface of normal unit vector $\mathbf{n}$ defined as in Figs.~\ref{Fig1} and \ref{Fig2} at $z=50$ \textmu m above a square pellet $20$ \textmu m long and $l=1$ \textmu m thick decomposed into 400 dipolar cubic volumes regularly distributed on a cubic network. The solid black square plotted on the figure delimits the contour of the system. (b) Temperature profile in the system calculated by inverse design to achieve the tilted emission shown in (a). (c) Another tilted emission pattern. (d) Temperature profile to achieve this target. The temperature of the external bath is $T_\mathrm{b}=300$ K.}\label{Fig6}
\end{figure}

In this approximation, the power density $p_\mathrm{dis}(\mathbf{r})$ needed to achieve the target emission can be calculated from the energy balance equation
\begin{equation}
p_\mathrm{dis}(\mathbf{r})=\nabla\cdot[\mathbf{S}(T(\mathbf{r}))-\lambda \nabla T(\mathbf{r})],
\label{energy_balance}
\end{equation}
where $\lambda$ sets the thermal conductivity of a solid, assuming here a diffusive heat-transport regime. The power density $p_\mathrm{dis}$ injected locally inside the system is plotted in Figs.~\ref{Fig4}(b) and \ref{Fig5}(b) for the same parallepipedic SiO$_2$ emitter as in Figs.~\ref{Fig1} and \ref{Fig2} in the cases of a normal and tilted emission, respectively. Here we have considered targets leading to temperature profiles with a weak transverse gradient so that Eq.~\ref{energy_balance} can be employed to calculate the required power density using the average temperature of the lower and upper dipolar layers. It is important to observe that the power injected or extracted at the system's edges dominates over that in the interior. This could present a practical challenge for the temperature control. Furthermore, strong thermal conduction may homogenize temperature gradients, thereby diminishing spatial control over emission directionality. In contrast, weak conduction helps maintain distinct thermal contrasts, preserving directional effects. Notice that incorporating convection or other interaction mechanisms between the solid and its environment could be straightforwardly achieved by adding additional terms to take into account these exchanges in Eq.(\ref{energy_balance}).

Finally, we illustrate the possibility to achieve a heat flux control using a two-dimensional system, a thin, $l=1$ \textmu m thick square pellet of SiO$_2$ whose surface is $20$ \textmu m long. In Figs.~\ref{Fig6}(a) and ~\ref{Fig6}(b) we show the targeted 2D Lorentzian emission patterns while in Figs.~\ref{Fig6}(b) and ~\ref{Fig6}(d) we plot the temperature profile solutions of the inverse problem with these targets. The comparison of these curves show that even a small local temperature disturbance around the equilibrium temperature $T_{\rm eq}=T_{\rm b}$ can significantly alter the direction of the emitted flux. Hence, a local temperature variation of few tenths of a degree can change the direction of emission by several degrees. This constitutes a technical challenge for the practical use of anisothermal sources.
From a practical standpoint, the temperature could be locally controlled using an embedded network of Peltier elements, which can heat or cool specific areas by adding or removing energy through the application of a bias voltage.

\subsection*{IV. CONCLUSION}
In conclusion, we have theoretically shown that controlling spatial temperature distributions within a solid can enable broadband, directional thermal emission. This approach utilizes anisothermal sources to overcome the inherent spatial incoherence of thermal radiation, providing a method for tailoring radiative heat flux at the microscale. By strategically designing temperature gradients, it is possible to achieve directional emission across a wide spectral range, without the need for complex nanophotonic structures. Notably, the directionality of thermal emission can be easily tuned by adjusting the temperature profile. Our findings open up new possibilities for advanced thermal management and energy conversion applications, where precise control of thermal radiation could improve efficiency and enable targeted heat transfer. More broadly, this work emphasizes the potential of anisothermal emitters to expand the design space for radiative energy systems by integrating microscale thermal gradients to control far-field emission. To finish it is important to emphasize that in the present work, we limited ourselves to the control of emission patterns radiated by polar materials. It would also be interesting to investigate the thermal behavior of metallic anisothermal sources. However, the DDA does not accurately account for eddy currents in metals. As a result, this method is not well-suited for such materials and a dedicated approach still needs to be developed to properly analyze metallic sources.

\subsection*{ACKNOWLEDGMENTS}
This work was supported by the French Agence Nationale de la Recherche (ANR), under grant ANR-21-CE30-0030 (NBODHEAT). F. H. acknowledges financial support by the Walter Benjamin Program of the Deutsche Forschungsgemeinschaft (eng. German Research Foundation) under project number 519479175.

\subsection*{DATA AVAILABILITY}

The data that support the findings in this article are available by contacting the authors.

\end{document}